\documentclass[12pt]{article}
\usepackage{graphicx}
\usepackage{float}
\usepackage{amsmath,color}
\title{\hfill {\small FZJ-IKP-TH-2005-9}\\
Partial-wave analysis of $\vec{p}\vec{p}\to pp\pi^\circ$
data}
\author{P. N. Deepak, J. Haidenbauer, and C. Hanhart \\
\small{Institut f\"ur Kernphysik (Theorie),
Forschungszentrum J\"ulich}\\
\small{52428 J\"ulich,
Germany}}
\date{}
\begin{document}
\maketitle
\begin{abstract}
We present a partial-wave analysis of the polarization data
for the reaction $\vec{p}\vec{p}\to pp\pi^\circ$, based solely on the
recent measurements at IUCF for this channel.
The fit leads to a $\chi^2$ per degree of freedom of 1.7.
Methods for an improved analysis are discussed.
We compare the extracted values to those from a meson exchange model.
\end{abstract}
\section{Introduction}
Understanding pion production in nucleon-nucleon ($NN$) collisions near 
threshold
is of high theoretical interest for various reasons.  As the
first strong inelasticity for the $NN$ system, its phenomenology is closely
linked to that of elastic $NN$ scattering (for recent reviews on the subject
of near-threshold pion production see Refs.~\cite{moskal,hanhart}).
In addition, as the pion is a
Goldstone-boson of the chiral symmetry of strong interactions, its dynamics
is strongly constrained by this symmetry (see
Ref.~\cite{ulfbible} and references therein).  Recently a scheme was 
discussed that
is said to lead to a convergent effective field theory even for large momentum
transfer reactions such as $NN\to 
NN\pi$~\cite{pwaves,withnorbert}. Confirmation of this claim is the
precondition for a successful analysis of the isospin violating pion
production reactions measured recently, namely the forward-backward asymmetry
in $pn\to d\pi^0$ \cite{allena} and the total cross section measurement
for $dd\to \alpha\pi^0$~\cite{stephen}.

A complete set of polarization observables for the reaction
$\vec p\vec p \to pp\pi^0$ was measured for the first time in 2001~\cite{meyer}.
Of the two existing advanced models of pion production in $NN$
collisions \cite{juelichdel,tamura} that include higher partial waves and
therefore allow predictions for polarization observables, only the model of the
J\"ulich group \cite{juelichdel,juelich} has been thoroughly
confronted with those data. Thereby it turned out that this model
failed to provide an overall satisfactory reproduction of these
polarization observables~\cite{meyer,juelichpol}.
On the other hand, the (less complete) data for $\vec p\vec p\to pn\pi^+$
\cite{daehnick} as well as those for
$\vec p\vec p\to d\pi^+$  \cite{barbara} were described very well by the 
same model.
So far the reason(s) for the short-coming of this phenomenological model to 
describe
the neutral pion production -- while being rather successful for the charged
pions -- is not yet understood\footnote{Note, however, that effective field
theory studies revealed many conceptual problems in this approach, as discussed
in \cite{hanhart}; it is up to now unclear how much impact those have
on the description of the observables.}.

\begin{table}[h!]
\begin{center}
\begin{tabular}{c|ccc}
\hline\hline
No.&Type&Our notation&Notation of Meyer et al.~\cite{meyer}\\
&&$T^{J}_{l_{q}(L_{p}s_f)j;Ls_{i}}$&$^{2s_i+1}L_J\to{}^{2s_f+1}L_{pj},l_q$\\[.1cm]\hline
1&Ss&$T^0_{0(00)0;11}$&$^3P_0\to{}^1S_0,s$\\[.1cm]
2&Ps&$T^0_{0(11)0;00}$&$^1S_0\to{}^3P_0,s$\\
3&&$T^2_{0(11)2;20}$&$^1D_2\to{}^3P_2,s$\\[.1cm]
4&Pp&$T^0_{1(11)1;11}$&$^3P_0\to{}^3P_1,p$\\
5&&$T^2_{1(11)1;11}$&$^3P_2\to{}^3P_1,p$\\
6&&$T^2_{1(11)2;11}$&$^3P_2\to{}^3P_2,p$\\
7&&$T^2_{1(11)1;31}$&$^3F_2\to{}^3P_1,p$\\
8&&$T^2_{1(11)2;31}$&$^3F_2\to{}^3P_2,p$\\
9&&$T^1_{1(11)0;11}$&$^3P_1\to{}^3P_0,p$\\
10&&$T^1_{1(11)1;11}$&$^3P_1\to{}^3P_1,p$\\
11&&$T^1_{1(11)2;11}$&$^3P_1\to{}^3P_2,p$\\
12&&$T^3_{1(11)2;31}$&$^3F_3\to{}^3P_2,p$\\[.1cm]
\hline
13&Sd&$T^2_{2(00)0;11}$&$^3P_2\to{}^1S_0,d$\\
14&&$T^2_{2(00)0;31}$&$^3F_2\to{}^1S_0,d$\\
15&Ds&$T^2_{0(20)2;11}$&$^3P_2\to{}^1D_2,s$\\
16&&$T^2_{0(20)2;31}$&$^3F_2\to{}^1D_2,s$\\\hline
\end{tabular}
\caption{Partial-wave amplitudes that could
contribute to $\vec{p}\vec{p}\to pp\pi^\circ$ near threshold.  Only 
contributions
arising from the first 12 amplitudes were considered in the present analysis.}
\label{table}
\end{center}
\end{table}

The presence of spin leads to contributions of many partial-wave amplitudes,
even close to threshold, which is the regime of interest here.  It is thus
difficult to draw any more concrete conclusion from a comparison of the model
results directly with the data.  It is well known that a partial--wave
analysis is an important intermediate step towards an understanding of
hadronic reactions: being in principle equivalent to the full data set, the
partial--wave amplitudes can be much more easily interpreted in terms of their
physics content.  As a consequence, a comparison of the theoretical results
with the partial--wave amplitudes is expected to reveal the
strengths/weaknesses of the theory much more clearly than a direct comparison
with the data.

In this paper we present a first step towards a full partial-wave
decomposition of the reaction $pp\to pp\pi^0$.  In our work we use as input
only data from the recent IUCF measurement \cite{meyer}.  However, as will be
stressed below, a combined analysis of both the production data and the data on
elastic scattering is mandatory for the future.  In \cite{meyer},
the various angular-dependent structures of the polarization observables were
fitted under particular assumptions on the partial wave content of the data as
well as on the energy dependence of some of the amplitudes. These assumptions
were necessitated by the limited statistical accuracy of the data. As we
use the extracted coefficients of Ref. \cite{meyer} as input for our fitting
procedure, we also have to make the same assumptions in our analysis.

This paper is organized as follows: in the next section we will describe the
theoretical formalism that allows to relate the observables to the 
partial-wave amplitudes.
In section 3 the method of extraction as well as that for
determining the uncertainties are explained.  Then, in section 4 we discuss 
the results
and compare them to those of a microscopic model 
\cite{juelichdel}.
The paper closes with a short summary and a discussion of further steps.

\section{Theoretical formalism}
The $\boldsymbol{T}$-matrix for $pp\to pp\pi^\circ$
may be expressed in the form~\cite{gr-pnd-msv}
\begin{align}
\boldsymbol{T}=\sum_{s_f,s_{i}=0}^1\ \sum_{\lambda=|s_i-s_f|}^{s_i+s_f} 
(S^{\lambda}(s_f,s_i)\cdot
T^{\lambda}(s_f,s_i)),
\label{cm}
\end{align}
where $s_i,s_f$ denote the initial and final channel-spins respectively.
We use the same notations as in~\cite{gr-msv}, where the irreducible
channel-spin transition operators $S^{\lambda}_{m_\lambda}(s_f,s_{i})$
of rank $\lambda$ are defined.  If in c.m., $\vec{p}_i,\vec{p}$ denote the
relative momenta of the two protons in the initial and final states and
$\vec{q}$ the momentum of the pion, the irreducible tensor reaction-amplitudes
$T^{\lambda}_{m_{\lambda}}(s_f,s_{i})$ in \eqref{cm}
can be expressed in the form~\cite{gr-pnd-msv}
\begin{align}
\label{irr-amps}
T^{\lambda}_{m_{\lambda}}(s_f,s_i)&=
\sum_{L_{p},L,l_{q}}\sum_{j,J,L_f}
(-1)^{L_f}
[j][L_f][J]^{2}[s_{f}]^{-1}
\begin{Bmatrix}
s_f&L_f&J\\
L&s_i&\lambda
\end{Bmatrix}
\begin{Bmatrix}
s_f&L_p&j\\
l_q&J&L_f
\end{Bmatrix}
\nonumber\\
&\times T^{J}_{l_{q}(L_{p}s_f)j;Ls_{i}}
((Y_{L_{p}}(\hat{p})\otimes Y_{l_{q}}(\hat{q}))^{L_f}\otimes
Y_{L}(\hat{p}_{i}))^{\lambda}_{m_{\lambda}}
\end{align}
to separate the energy and angular dependence
of the amplitudes.  In Eq. \eqref{irr-amps}, we use
the short-hand notation $[j]=\sqrt{2j+1}$ and $(T_1 \otimes T_2)^L_m$
indicates the  coupling
of the two irreducible tensors $T_1$ and $T_2$ to total angular momentum $L$
with
projection $m$.
The partial-wave amplitudes
$T^{J}_{l_{q}(L_{p}s_f)j;Ls_{i}}$
are functions of both the c.m. energy $E_\text{cm}$ and $\epsilon$,
the relative kinetic energy of the nucleon pair in the final
state (in contrast to a two body reaction, where the partial-wave amplitudes
are characterized by a single energy variable).

If $\vec{P},\,\vec{Q}$ denote respectively the beam and target polarizations,
the differential cross section in a double-polarized
experiment may be written as~\cite{gr-pnd-msv}
\begin{equation}
{{\text{d}\sigma}\over {\text{d}\Omega_p\text{d}\Omega_q\text{d}\epsilon}}=
\frac{1}{4}\sum _{k_{1},k_{2}=0}^{1}
\ \sum _{k=|k_{1}-k_{2}|}^{k_{1}+k_{2}}
( ( P^{k_{1}}\otimes Q^{k_{2}})^{k}
\cdot B^{k}(k_{1},k_{2})),
\label{dcs}
\end{equation}
in terms of the irreducible tensors
\begin{align}
&B^{k}_{\nu}(k_{1},k_{2})=2(-1)^{k_{1}+k_{2}}[k_{1}][k_{2}]\sum
_{s_{f}=0}^{1}(2s_{f}+1)\sum _{s_{i},s_{i}'=0}^{1}
\sum _{\lambda,\lambda'}
(-1)^{s_{i}'+s_{f}}
[s_{i}][s_{i}'][\lambda][\lambda']\nonumber\\
&\times
\begin{Bmatrix}
s_i'&s_i&k\\
\lambda&\lambda'&s_f
\end{Bmatrix}
\ \begin{Bmatrix}
\textstyle{\frac{1}{2}}&\textstyle{\frac{1}{2}}&s_i\\[.2cm]
\textstyle{\frac{1}{2}}&\textstyle{\frac{1}{2}}&s_i'\\[.2cm]
k_1&k_2&k
\end{Bmatrix}\,( T^{\lambda}(s_{f},s_{i})\otimes
T^{\dagger ^{\lambda'}}(s_{f},s_{i}'))^{k}_{\nu},
\label{bi}
\end{align}
which are bilinear in the irreducible
tensor amplitudes
$T^{\lambda}_{m_{\lambda}}(s_{f},s_{i})$ whose complex
conjugates
$T^{\lambda}_{m_{\lambda}}(s_{f},s_{i})^{*}$ define
$T^{\dagger ^{\lambda}}_{m_{\lambda}}(s_{f},s_{i})=
(-1)^{m_{\lambda}}T^{\lambda}_{-m_{\lambda}}(s_{f},s_{i})^{*}
$.  If
\begin{align}
\label{unpol-dcs}
\sigma_0(\xi)=\frac{1}{4}B^0_0(0,0)
\end{align}
denotes the unpolarized differential cross section with
$\xi$ collectively standing for $\{\hat{p},\hat{q},\epsilon\}$, the 
$B^k_\nu(k_1,k_2)$
are related to the independent (Cartesian) spin-observables $A_{ij}(\xi)$,
defined in \cite{meyer}, through
\begin{subequations}
\begin{align}
\label{obsfirst}
\sigma_0(\xi)\,A_{y0}(\xi)&={\textstyle{\frac{-1}{2\sqrt{2}}}}\ 
\Im(B^1_1(1,0));\\
\sigma_0(\xi)\,A_{xz}(\xi)&={\textstyle{\frac{1}{4}}}\ 
[\Re(B^1_1(1,1)-B^2_1(1,1))];\\
\sigma_0(\xi)\,A_{\Sigma}(\xi)&={\textstyle{\frac{-1}{2\sqrt{3}}}}[B^0_0(1,1)+
{\textstyle{\frac{1}{\sqrt{2}}}}B^2_0(1,1)];\\
\sigma_0(\xi)\,A_{zz}(\xi)&={\textstyle{\frac{-1}{4\sqrt{3}}}}[B^0_0(1,1)-\sqrt{2}B^2_0(1,1)];\\
\sigma_0(\xi)\,A_{\Delta}(\xi)&={\textstyle{\frac{1}{2}}}\ \Re(B^2_2(1,1));\\
\sigma_0(\xi)\,A_{z0}(\xi)&={\textstyle{\frac{1}{4}}}\ B^1_0(1,0);\\
\label{obslast}
\sigma_0(\xi)\,A_{\Xi}(\xi)&={\textstyle{\frac{-1}{2\sqrt{2}}}}\ 
\Im(B^1_0(1,1).
\end{align}
\end{subequations}
\begin{figure}[t]
\begin{center}
\includegraphics[scale=.6,keepaspectratio,angle=270]{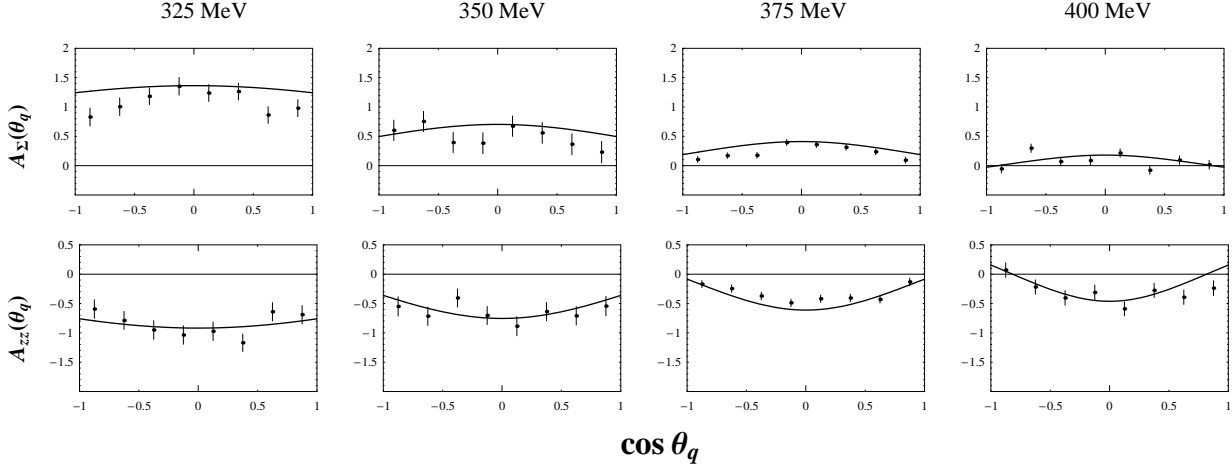}
\caption{The observables $A_\Sigma(\theta_q)$ and $A_{zz}(\theta_q)$ as a function of the pion angle
at several bombarding energies.  The data and the nomenclature  for the observables
are taken from Meyer et al.~\cite{meyer}. The
solid lines represent our results.}
\label{obs1}
\end{center}
\end{figure}

\section{Extraction of partial-wave amplitudes}
A priori, a set of 16 partial-wave amplitudes can be expected
to contribute to the reaction.  We list them in Table~\ref{table},
explicitly both in our notation and the notation of Meyer et
al.~\cite{meyer}.  But we consider only contributions from the first 12
amplitudes since final states with orbital angular momentum
greater than 1 were ignored in the
analysis of Meyer et al.~\cite{meyer}.
Thus there are 24 real unknowns (12 complex amplitudes) to be determined.
However, as overall phases are unobservable and $s_f=0$
and $s_f=1$ $NN$-final-states do not mix with each other in any
of the spin-observables measured in~\cite{meyer} (final state polarizations
were not measured), we have the freedom to choose the first
two amplitudes to be real.  This leaves 22 real
numbers to be determined.
Equations (11a)-(11h) of \cite{meyer} represent the
general angular dependence of
$\sigma_0(\xi)$ and $\sigma_0(\xi)A_{ij}(\xi)$
in terms of the real coefficients $E,\,F_k,\,G^{ij}_k,\,
H^{ij}_k,\,I,\,K,\,I^{ij}$ and $K^{ij}$.   The quantities
$E,\,F_k,\,G^{ij}_k$ and $H^{ij}_k$ denote
the weighted sums of bilinears in the partial-wave
amplitudes corresponding respectively to $(\text{Ss})^2,\,
(\text{Ps})^2,\,(\text{PsPp})$ and $(\text{Pp})^2$ interference terms, while
$I,\,I^{ij}$ and $K,\,K^{ij}$ represent respectively the
contribution of $(\text{SsSd})$ and $(\text{SsDs})$ interference terms, 
which were
ignored in the analysis of \cite{meyer} and therefore also here.
Using Eqs. \eqref{bi} and \eqref{irr-amps}, we obtain
explicit expressions for all the observables
in \eqref{obsfirst}-\eqref{obslast} including the
unpolarized differential cross section, defined in \eqref{unpol-dcs},
in terms of the first 12 partial-wave amplitudes listed in Table~\ref{table}.
These expressions when compared with Eqs. (11a)-(11h) of
\cite{meyer}, allow us to obtain explicitly the
partial-wave decomposition of the coefficients
$E,\,F_k,\,G^{ij}_k$ and $H^{ij}_k$.

\begin{figure}[t]
\begin{center}
\includegraphics[scale=.6,keepaspectratio,angle=270]{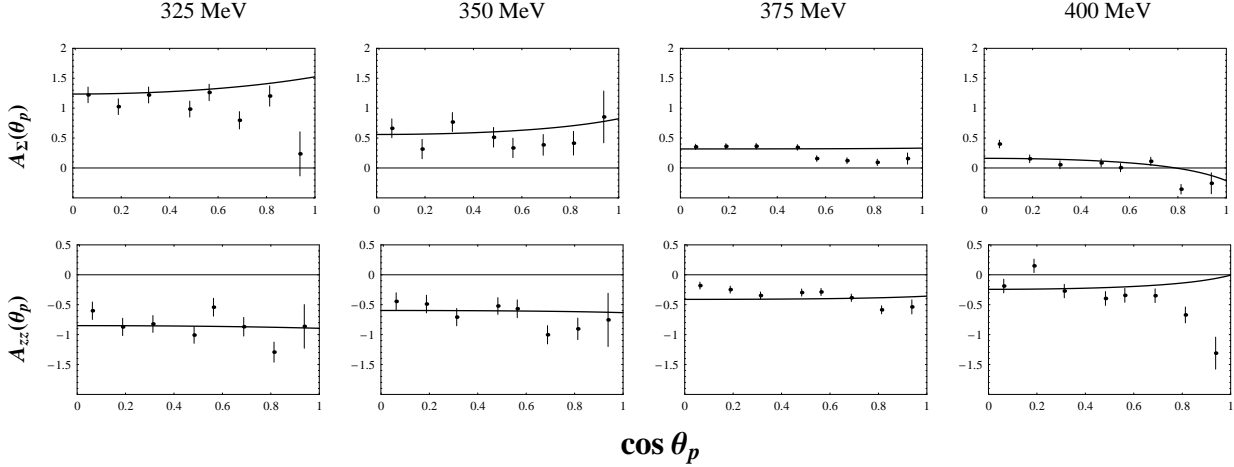}
\caption{The observables $A_\Sigma(\theta_p)$ and $A_{zz}(\theta_p)$ as a function of the proton angle
at several bombarding energies.}
\label{obs2}
\end{center}
\end{figure}

Note that all values given in Table IV of
\cite{meyer} for the various coefficients
are integrated with respect to the outgoing two nucleon energy, $\epsilon$.
Thus, they are expressed as weighted sums of numerous
$\mathcal{B}_{\kappa\kappa'}$, bilinear
in the partial-wave amplitudes, as
(we use non--relativistic kinematics)
\begin{align}
\label{iifdef}
\mathcal{B}_{\kappa\kappa'}(E_\text{cm})=
\int^{\epsilon_\text{max}}_0 
T_\kappa(E_\text{cm},\epsilon)\,T_{\kappa'}^*(E_\text{cm},
\epsilon)\,q(E_\text{cm},\epsilon)
\,p(\epsilon)\,\text{d}\epsilon \ ,
\end{align}
where $T_\kappa,\,\kappa=2,\ldots,12$, denotes the $\kappa^\text{th}$ 
partial-wave
amplitude listed in Table~\ref{table} (for example, 
$T_5(E_\text{cm},\epsilon)\equiv
T^2_{1(11)1;11}$), $p(\epsilon)=\sqrt{M_N\epsilon}$ and
$q(E_\text{cm},\epsilon)=\sqrt{2\mu(E_\text{cm}-2M_N-m_\pi-\epsilon)}$, 
with the reduced mass
of the outgoing three body system $\mu=2m_\pi M_N/(m_\pi+2M_N)$, where
$M_N$ and $m_\pi$ are the nucleon and pion masses, respectively.
Thus, to proceed further we need
to make an assumption regarding the $\epsilon$-dependence of the
$T_\kappa$.  In the present first analysis of the IUCF data we use the most 
naive ansatz possible:
we assume that the entire energy dependence of the amplitudes stems solely 
from the
centrifugal barrier. This gives
\begin{equation}
T_\kappa(E_\text{cm},\epsilon)\propto q(E_\text{cm},
\epsilon)^{l_q(\kappa)}p(\epsilon)^{L_p(\kappa)} \ ,
\label{rosen}
\end{equation}
which should hold as long as the outgoing momenta are small compared to the
inverse of the production radius \cite{rosenfeld} and the effects of the
final state interaction (FSI) are negligible. (This is obviously wrong for the
$NN$ $S$--waves; they are discussed separately below.)
Note, the same assumption
was also used in the fitting procedure of Ref. \cite{meyer} in order to
determine some of the coefficients $E,\,F_k,\,G^{ij}_k$ and $H^{ij}_k$ from 
the
data, for the statistical accuracy of the data did not allow for a
separate fit of these coefficients at each energy.
 From the ansatz of Eq. \eqref{rosen}, one easily derives
\begin{align}
\label{para}
\mathcal{B}_{\kappa\kappa'}(E_\text{cm})=
z_\kappa\,z_{\kappa'}^*\ 
\eta^{l_q(\kappa)+l_q(\kappa')+L_p(\kappa)+L_p(\kappa')+4} \ .
\end{align}
Thus, we find the energy dependence of the 
$\mathcal{B}_{\kappa\kappa'},\,\kappa,\kappa'=2,\ldots,12$,
to be of the form $\eta^x$, with $x$ equal
to 6,7 and 8 for the PsPs, PsPp and PpPp interference terms, respectively.
Here $\eta=q_{\text{max}}/m_\pi$ with $q_{\text{max}}$ being
the largest possible value of pion momentum for a given incident energy.
By assumption, $z_\kappa,\,\kappa=2,\ldots,12$ in \eqref{para}
are energy-independent complex quantities to be determined from the data.

Since the transition amplitude with the Ss final-state does not interfere
with any of the other partial waves and since its
FSI does not show a power law behavior \cite{mus}, we parameterize it as
\begin{align}
\mathcal{B}_{11}(E_\text{cm})=
\int^{\epsilon_{\text{max}}}_0 |T_1(E_\text{cm},\epsilon)|^2
\,q(E_\text{cm},\epsilon)\,p(\epsilon)\,\text{d}\epsilon=
|z_1|^2
\end{align}
and extract it at each of the four bombarding energies individually;
$\mathcal{B}_{11}(E_\text{cm})$ is directly proportional to the bilinear 
coefficient
$E$ in Table IV of Ref. \cite{meyer}.

\begin{figure}[t]
\begin{center}
\includegraphics[scale=.57,keepaspectratio,angle=270]{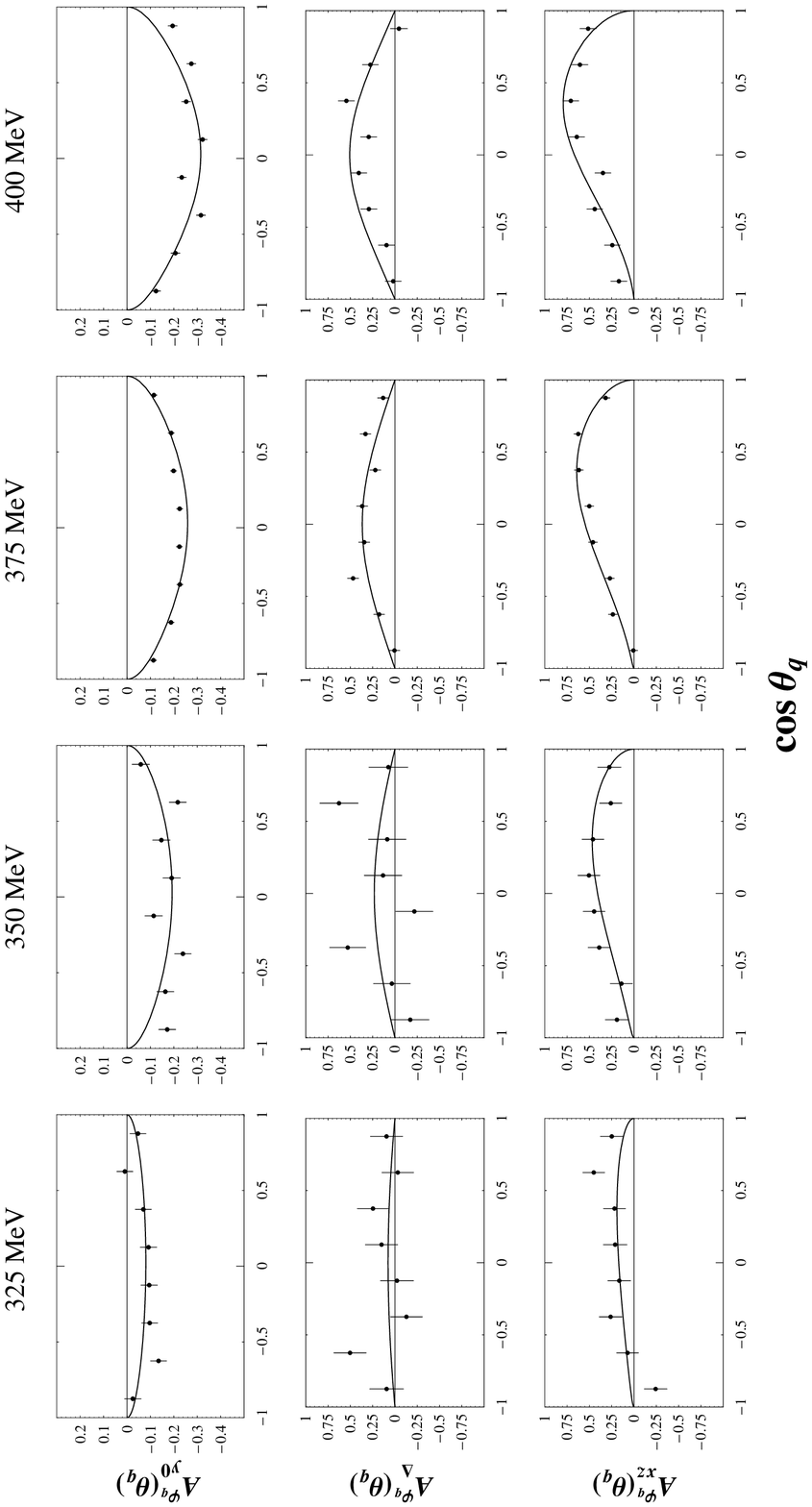}
\caption{The observables $A^{\varphi_q}_{y0}(\theta_q)$, 
$A_{xz}^{\varphi_q}(\theta_q)$
and $A^{\varphi_q}_{\Delta}(\theta_q)$ as a function of the pion angle
at several bombarding energies.}
\label{obs3}
\end{center}
\end{figure}

\begin{figure}[t]
\begin{center}
\includegraphics[scale=.57,keepaspectratio,angle=270]{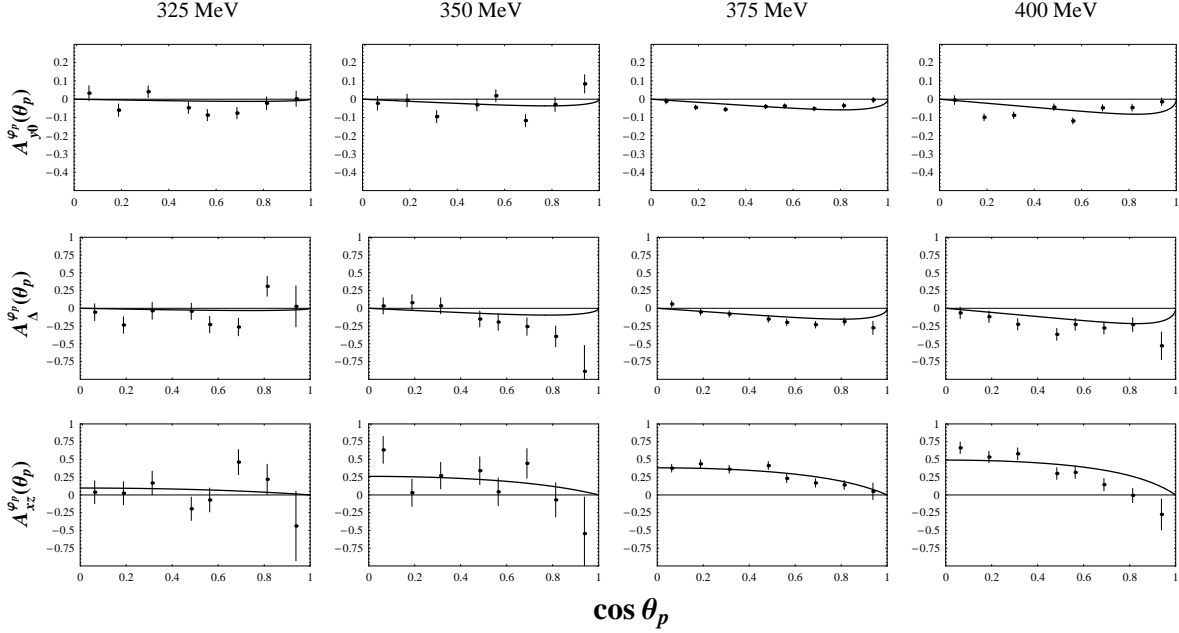}
\caption{The observables $A^{\varphi_p}_{y0}(\theta_p)$, 
$A_{xz}^{\varphi_p}(\theta_p)$
and $A^{\varphi_p}_{\Delta}(\theta_p)$ as a function of the proton angle
at several bombarding energies.}
\label{obs4}
\end{center}
\end{figure}

The values of the coefficients $H_1^{00},\,H^{zz}_1,\,F_2,\,H^\Sigma_2,
\,H^{zz}_2,\,G^{z0}_1,\,G^\Xi_1,\,H^{z0}_1,\,H^{z0}_2,
\,H^\Sigma_4,\,H^\Sigma_5,\,H^{zz}_4$ and $H^{zz}_5$ were determined by 
Meyer et al.~\cite{meyer}
by assuming their energy dependence to be of the form given in Eq. 
\eqref{para}.
Therefore, in order to be consistent with~\cite{meyer}, we do the
following.  We plot each of these 13 coefficients as a function of their
appropriate $\eta$-dependence and extract the slope and the corresponding
error, using the values and errors in Table IV of \cite{meyer}.
For example, $(\sigma_{tot}(\eta)F_2(\eta))/(8\pi^2)$ is plotted
against $\eta^6$.  The values and errors so obtained were then used in our 
fitting procedure.
This gives us a set of 13 equations for the amplitudes $z_\kappa$, 
$\kappa=2,\ldots,12$.
The remaining 26 coefficients, viz., $F_1,\,H^\Sigma_0,\,H^{zz}_0$ and the 
23 coefficients
from $G^{y0}_1$ to $H^\Delta_5$, in Table IV of \cite{meyer} lead to
$26\times 4=104$ equations for the amplitudes
$z_\kappa$, $\kappa=2,\ldots,12$, as they were extracted by \cite{meyer}
without any assumption about their energy dependence.
Both the values and errors for these 104 coefficients,
taken from Table IV of \cite{meyer}, were multiplied by 
$(\sigma_{tot}(\eta)/(8\pi^2))$
for consistency.
In total, we have 117 equations with 21 real unknowns
(since $z_2$ is assumed to be real).
In addition, the coefficient $E$ depends only on $z_1$. Since this amplitude
is assumed to be real and we know the explicit dependence of $E$ on $z_1$,
it is directly determined (up to a sign).  The uncertainty in $z_1$, is
determined by that in $E$.

\begin{figure}[t]
\begin{center}
\includegraphics[scale=.5,keepaspectratio,angle=270]{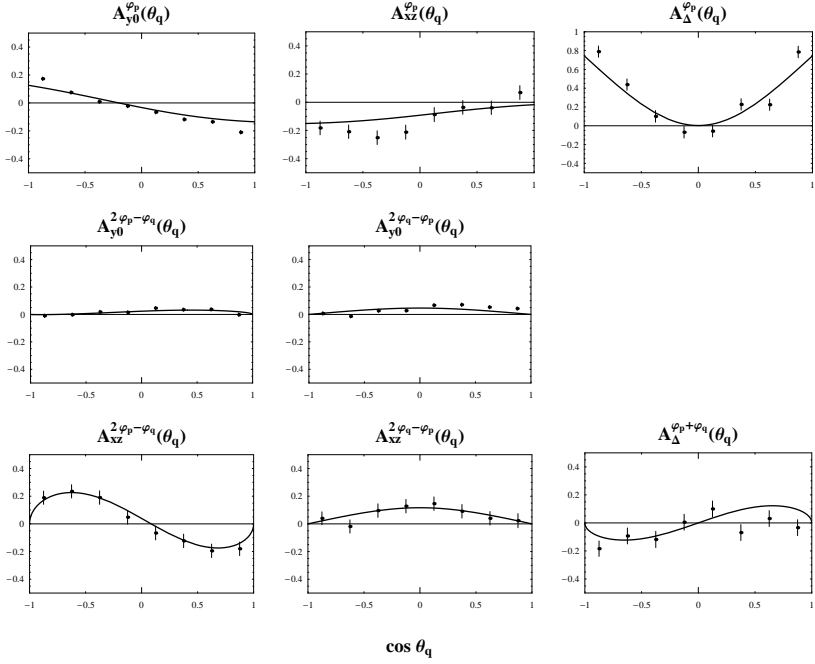}
\caption{Additional observables at a bombarding energy of 375 MeV
as a function of the pion angle.}
\label{obs5}
\end{center}
\end{figure}

\begin{figure}[h!]
\begin{center}
\includegraphics[scale=.5,keepaspectratio]{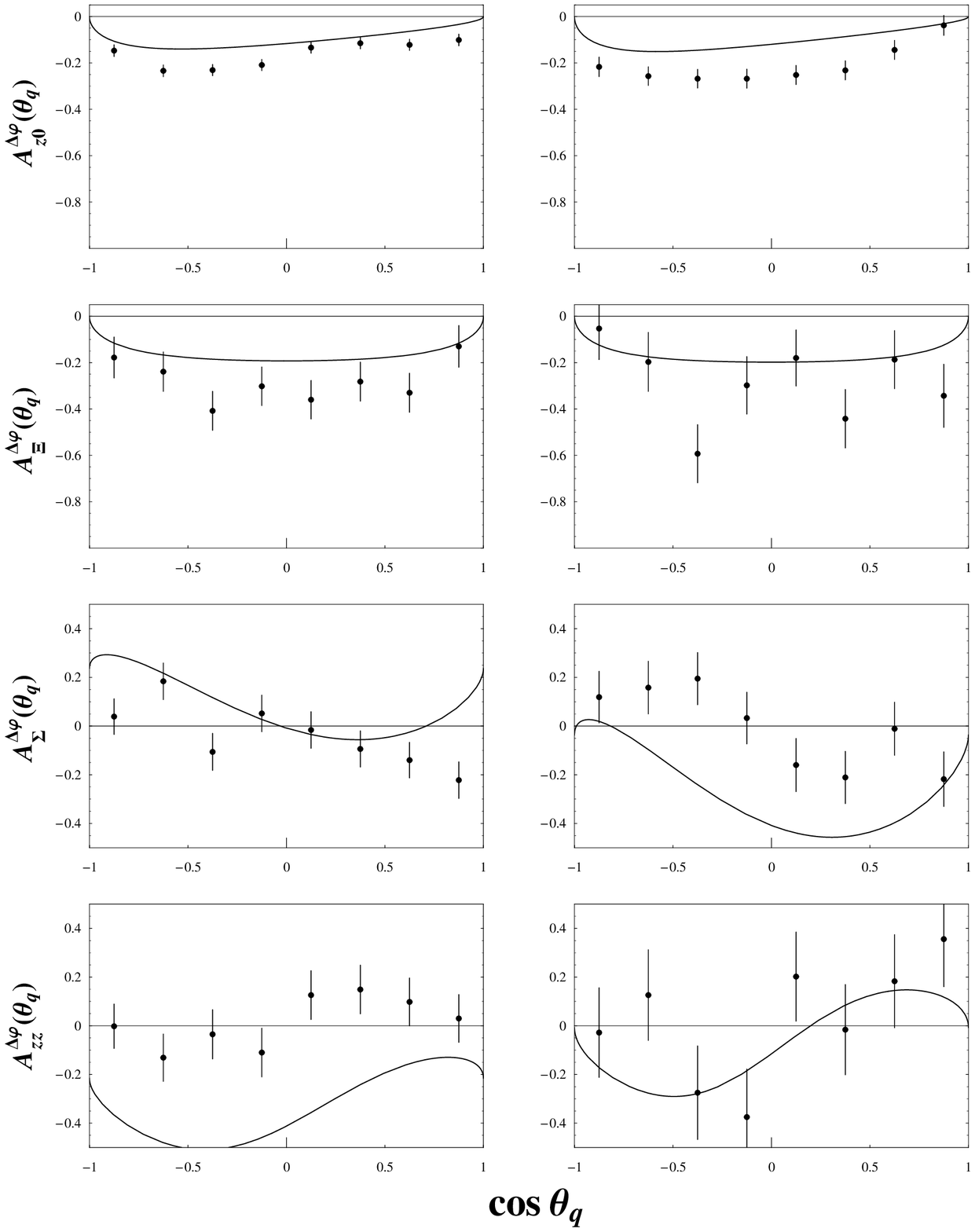}
\caption{Observables that depend on $\Delta\varphi_q$,
plotted as a function of the angle $\theta_q$ at the bombarding
energies of 375 and 400 MeV, at which the measurement of Ref. \cite{meyer}
has the best statistics.
}
\label{obs6}
\end{center}
\end{figure}

This nonlinear overdetermined system of 117 equations
can only have approximate solutions, which were obtained by 
$\chi^2$-minimization using the software {\textit{Mathematica}}.
The resulting $\chi^2$ per degree of freedom was 1.7.
This value was the best that we could obtain
after using all the four methods
of minimization available with {\textit{Mathematica}}, viz., differential 
evolution,
Nelder-Mead, random search
and simulated annealing.
We performed various checks (different starting vectors; setting individual
amplitudes to zero) to further support that the minimum is indeed a total 
minimum.
In Figs. \ref{obs1}-\ref{obs6}, to illustrate the quality of the fit,
we compare our results to some of the observables measured in \cite{meyer}.

The uncertainties in the $z_\kappa$, $\kappa=2,\ldots,12$, were determined as 
follows.
Let $\boldsymbol{a}$ denote the vector whose 21 components are the real and
imaginary parts of the amplitudes $z_i$. Then the uncertainty (standard 
deviation) in
any of the components, say $a_j$, denoted by $\sigma_{a_j}$, is obtained
through~\cite{burrell}
\begin{align}
\sigma_{a_j}^2=\sum_{i=1}^{117}\left(\sum_{l=1}^{21}
\frac{1}{\sigma_i}\left(\epsilon_{jl}
\frac{\partial}{\partial a_l}f_i(\boldsymbol{a})\right)_
{\boldsymbol{a}=\boldsymbol{a_0}}\right)^2 \ .
\end{align}
Here $f_i(\boldsymbol{a})$ stands for the explicit functional form of the
bilinear coefficients listed in Table IV of \cite{meyer}, in terms of the
$z_\kappa$.  $\sigma_i$ are the corresponding errors in these bilinear
coefficients, taken from the same Table and $\epsilon_{jl}$ is the $(j,l)$--th
element of the error (covariance) matrix $\boldsymbol{\epsilon}$ defined as
the inverse of the curvature matrix $\boldsymbol{\alpha}$ whose elements are
given by
\begin{align}
\alpha_{jl}\equiv\frac{1}{2}\frac{\partial^2\chi^2}
{\partial a_j\partial a_l} \ .
\end{align}
Here $\boldsymbol{a_0}$ is the
value of $\boldsymbol{a}$ for which the value of $\chi^2$ is at its minimum.

\section{Results and discussion}

\begin{figure}[t]
\begin{center}
\includegraphics[width=.42\textwidth]{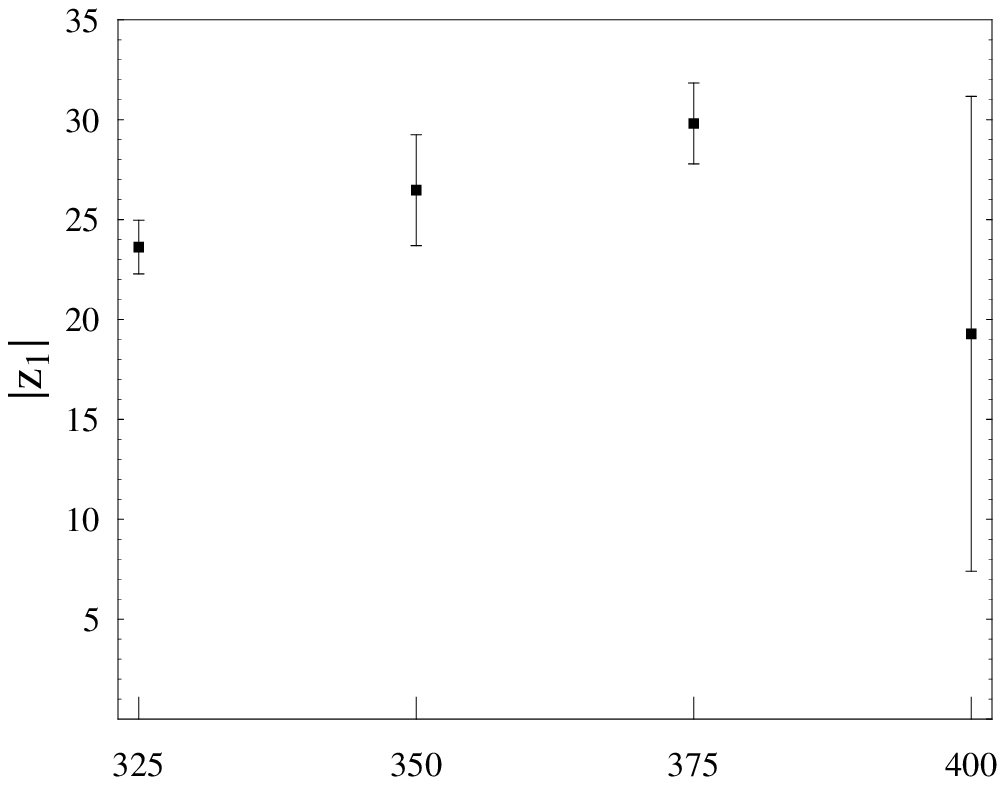}
\includegraphics[width=.42\textwidth]{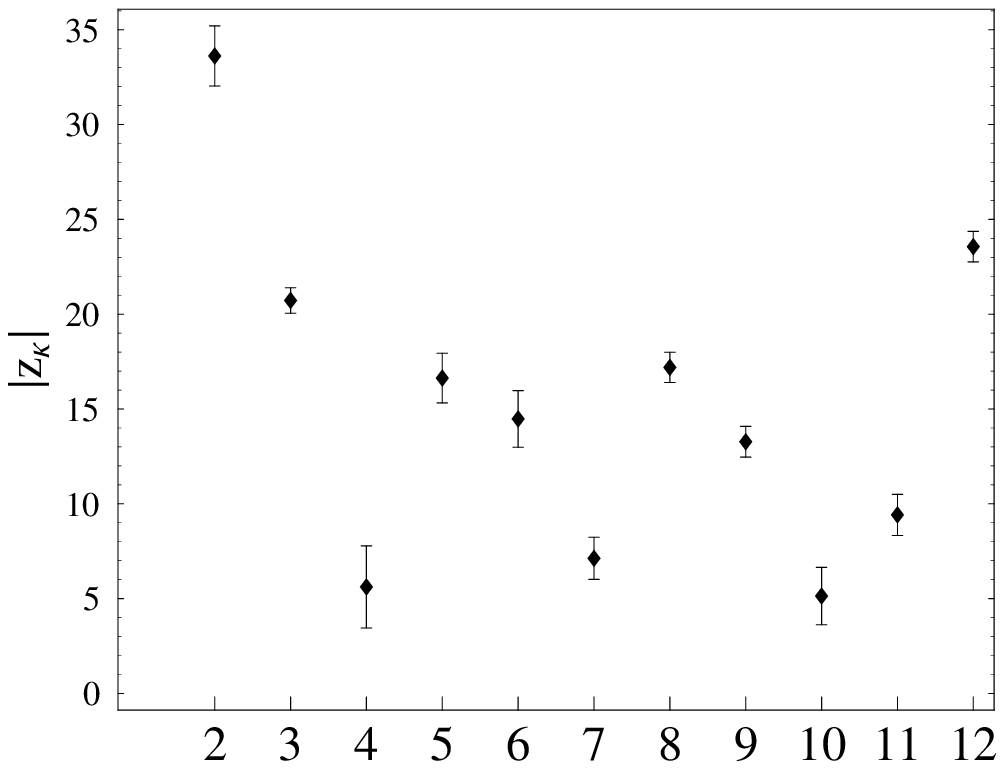}
\caption{\label{z1}
Left panel: Moduli (in $(\mu\text{b})^{1/2}$) of $z_1$
for bombarding energies of 325, 350, 375, and 400 MeV.
Right panel: Moduli (in $(\mu\text{b})^{1/2}$) of the $z_\kappa,\ 
\kappa=2,\ldots,12$.
}
\end{center}
\end{figure}

\begin{figure}[t]
\begin{center}
\includegraphics[width=.4\textwidth]{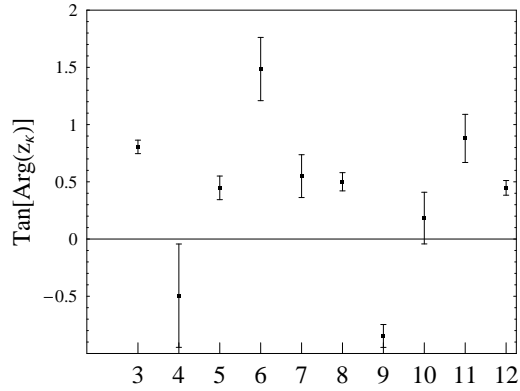}
\caption{\label{phz}
Tangent of the arguments of the $z_\kappa,\ \kappa=3,\ldots,12$.}
\end{center}
\end{figure}

In Figs. \ref{z1}-\ref{phz}, the values for the $z_\kappa$, 
$\kappa=1,\ldots,12$, as
determined in the fit are plotted.  The uncertainties
quoted in these figures for $|z_\kappa|$, $\kappa=1,\ldots,12$ and 
$\tan(\text{Arg}(z_\kappa))$,
$\kappa=3,\ldots,12$, were determined by propagating the errors obtained 
for the
real and imaginary parts of the $z_\kappa$, $\kappa= 1,\ldots,12$,
in the standard way~\cite{bevington}.
It is striking that the amplitude $z_2$, corresponding to the transition $^1S_0\to
^3P_0s$, is the largest.

\begin{figure}[h]
\begin{center}
\includegraphics[width=.6\textwidth]{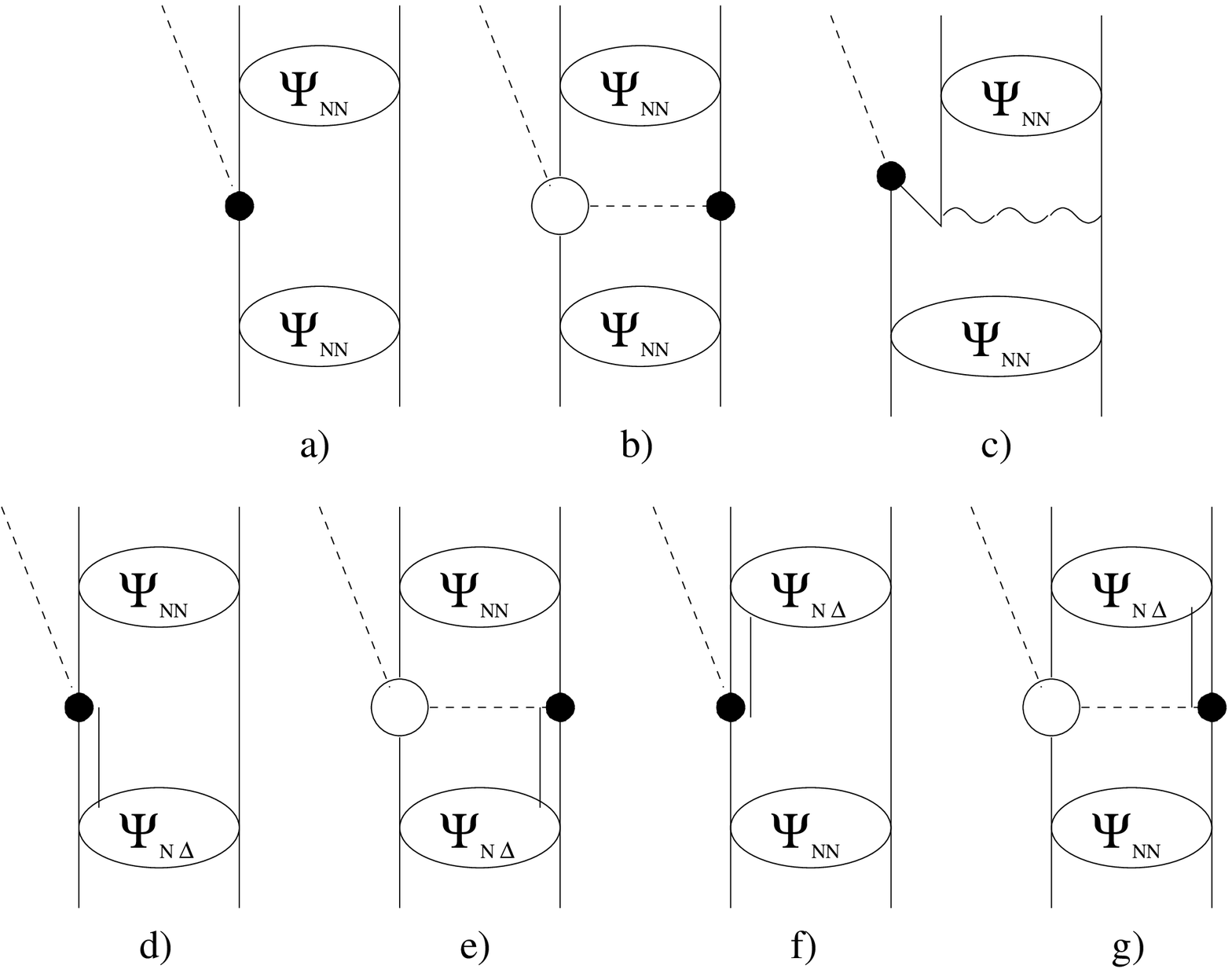}
\caption{\label{beitraege}
Pion production mechanisms taken into account in the model
of Ref. \protect\cite{juelichdel}:
(a) direct production; (b) pion rescattering; (c) contributions from pair
diagrams; (d) to (g) production involving the excitation of
the $\Delta (1232)$ resonance, depicted by the double line.
In the diagrams the pion (nucleon) is shown as a dashed (solid) line. }
\end{center}
\end{figure}

We now turn to a comparison of the extracted $z_\kappa$ with the predictions
of the microscopic model of the J\"ulich group.  For a detailed description
of the model we refer the reader to Refs. \cite{juelichdel,juelichpol}.
Here we only want to summarize its salient features.
In the J\"ulich model all standard pion-production mechanisms
(direct production (Fig.~\ref{beitraege}a),
pion rescattering (Fig.~\ref{beitraege}b),
contributions from pair diagrams (Fig.~\ref{beitraege}c))
are considered. In addition, production mechanisms involving the
excitation of the $\Delta (1232)$ resonance
(cf. Fig.~\ref{beitraege}d--g) are taken into account explicitly.
All $NN$ partial waves up to orbital angular momenta $L_p = 2$, and
all states with relative orbital angular momentum $l_q \leq 2$ between
the $NN$ system and the pion are considered in the final state.
Furthermore all $\pi N$ partial waves up to orbital angular momenta
$L_{\pi N} = 1$ are included in calculating the rescattering diagrams
in Fig.~\ref{beitraege}b,e and g. Thus, this model includes not only
s-wave pion rescattering but also contributions from p-wave rescattering.

The reaction $NN \to NN\pi$ is treated in a distorted wave born approximation,
in the standard fashion. The actual calculations are carried out
in momentum space.  For the distortions in the initial and final $NN$ states,
a coupled channel ($NN$, $N\Delta$, $\Delta\Delta$) model
is employed \cite{HHJ} that
treats the nucleon and the $\Delta$ degrees of freedom on equal footing.
Thus, the $NN \leftrightarrow N\Delta$ transition amplitudes
and the $NN$ T--matrices that enter in the evaluation of the pion
production diagrams in Fig. \ref{beitraege} are consistent solutions of
the same (coupled--channel) Lippmann--Schwinger--like equation.

By taking the partial-wave amplitudes $T_\kappa(E_\text{cm},\epsilon)$ predicted
by the model, it is straightforward to extract the moduli of the $z_\kappa$
from the model through their definition in Eqs. (\ref{iifdef})
and (\ref{para}):
$$
|z_\kappa^{model}| = \eta^{-l_q(\kappa)-L_p(\kappa)-2}\sqrt{
\mathcal{B}_{\kappa\kappa}(E_\text{cm})} \ .
$$
Since 
the phases of the bilinears $\mathcal{B}_{\kappa\kappa'}$ calculated from the
model 
are not in all cases consistent  with the factorization
used in Eq. (\ref{para}), they can not be compared easily with
those extracted from the data.

\begin{figure}[h!]
\begin{center}
\includegraphics[width=.7\textwidth]{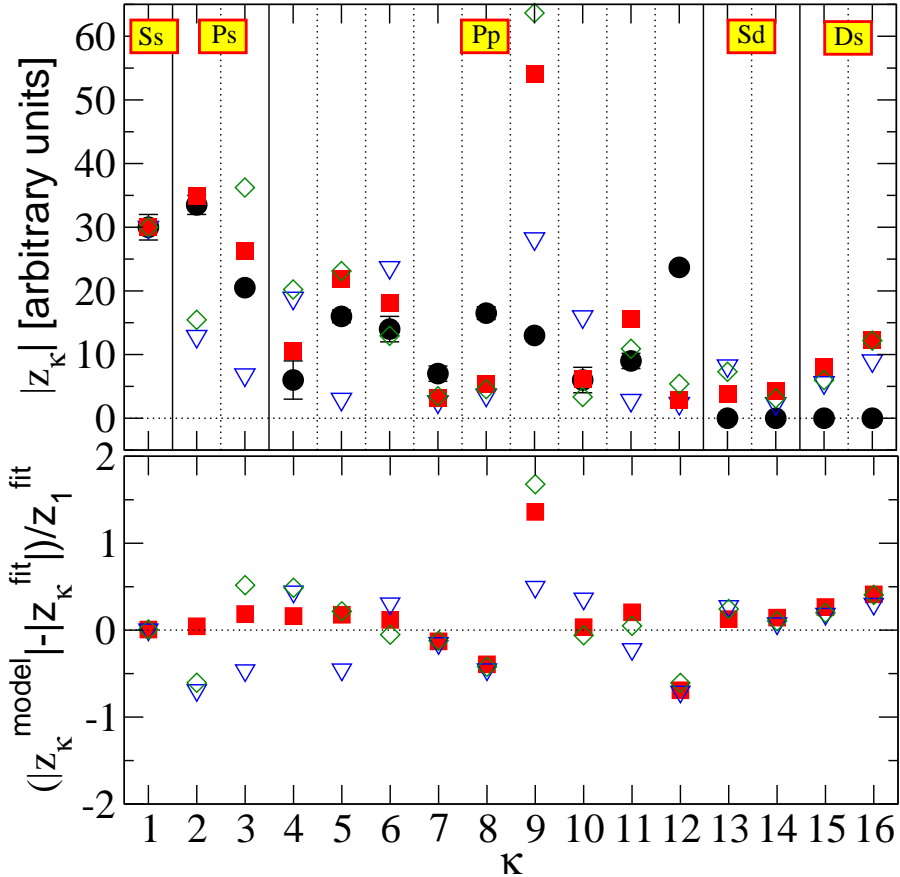}
\caption{\label{modelcomp}
Comparison of the extracted $|z_\kappa|$ (filled circles) to the 
corresponding quantities
predicted by the microscopic model of Ref. \cite{juelichdel} (filled squares).
The opaque triangles show results where the $\Delta$ contributions of the 
model
were omitted completely whereas the opaque diamonds represent results of a 
calculation
where only the $\Delta$ contributions after pion emission were omitted.
In the upper panel the results
for the various $|z_\kappa|$ are shown, normalized with respect to
our extracted value for $|z_1|$ at 375 MeV, while in the
lower panel, the relative deviations of the model calculation from the
extracted values are shown.
}
\end{center}
\end{figure}

The $|z_\kappa|$ predicted by the model are compared with the results of
the partial-wave analysis in Fig.~\ref{modelcomp}.
In the upper part of the graph we compare the moduli of the
$z_\kappa$ of the model with those obtained by our partial-wave analysis,
while in the lower part, the deviation of the model predictions from the
analysis are presented.  Note that the model results
are normalized in such a way that $|z_1|$ for a bombarding
energy of 375 MeV (i.e. the one
corresponding to the $^3P_0\to^1S_0s$ transition) coincides with
the corresponding extracted value.  This is done in order to facilitate the
comparison between the various other $|z_\kappa|$.

It is evident from Fig.~\ref{modelcomp} that the microscopic model of
Refs. \cite{juelichdel,juelichpol} yields a rather impressive overall 
description
of the various partial-wave amplitudes, cf. the filled squares and
circles.  This is particularly remarkable
because one has to keep in mind that practically all parameters of the
model were fixed by other reactions (elastic-$NN$-scattering and $\pi N$
scattering)\footnote{The only free parameter  was fixed
to the total cross section for $pp\to pp\pi^0$ at low energies
\cite{juelich}.}.
In fact the majority of the partial-wave amplitudes is reproduced
even quantitatively (if one takes into account the error bars of the
partial-wave analysis).  The only serious discrepancy occurs
in the amplitude $z_9$ ($^3P_1\to^3S_1p$) and to a lesser extent
also in $z_{12}$ ($^3F_3\to^3P_0p$).  The reason for the short-coming
in the model prediction for these $|z_\kappa|$ and the connection
with its dynamical ingredients  needs to be explored
in the future.

Fig.~\ref{modelcomp} suggests also considerable deviations in 
$z_{13}$, $z_{15}$  and especially in $z_{16}$. However, these 
$z_\kappa$ correspond to partial waves with 
$NN$ $D$-waves or pion $d$-waves in the final state
whose contribution had been set to zero in the analysis of
Meyer et al. \cite{meyer}---as well as in ours---as already mentioned above.
Thus, the predictions of the microscopic model can be seen as an
 indication that those amplitudes may not be negligible
and therefore should be taken into account in
any future analysis of the reaction $pp\to pp\pi^0$.
Since in the present analysis, the neglected contributions of
the $D$-wave and $d$-wave amplitudes are presumably mimicked by other
partial-wave amplitudes, a more complete
partial-wave analysis could yield results that are even closer
to the model prediction than for the case considered in the
present paper.

While a well-founded theoretical interpretation of the obtained
partial-wave amplitudes calls for a thorough investigation, e.g.
within the framework of effective field theory, as advocated in
Refs.~\cite{hanhart,pwaves,withnorbert},
the presented analysis allows already to shed light on
the role of the $\Delta$ (1232) resonance for $\pi^0$
production.
The importance of the $\Delta$ isobar for the reaction
$NN\to NN\pi$ was already pointed out in Ref.
\cite{juelichdel}.  The present partial-wave analysis allows to confirm
that aspect nicely in a quantitative and transparent way.
The results of the model of Ref.~\cite{juelichdel,juelichpol} after
omitting contributions involving $\Delta$ degrees of freedom are
shown by the triangles in Fig.~\ref{modelcomp}.  The
corresponding predictions clearly fall short in describing
the amplitudes of the partial-wave analysis.  In particular,
even the qualitative trend in the magnitude of the amplitude
is not reproduced.

Further insight can be gained by taking into account only
those $NN \to N\Delta$ transitions that occur before the
pion emission (Fig.~\ref{beitraege}d and e).  The corresponding predictions 
for the $|z_\kappa|$ are shown by the open diamonds in Fig.~\ref{modelcomp}.
For almost all partial waves this part provides the dominant $\Delta$ 
effect, as expected, since the energy of the incoming $NN$ system is not too far
away from the nominal $N\Delta$ threshold. The energy in the
outgoing $NN$ system, on the other hand, is much smaller and therefore the
excitation of the $\Delta$(1232) is expected to be of much
less significance. However, to achieve also a quantitative agreement with the
extracted $|z_\kappa|$, the $\Delta$ excitation in the final state (after 
pion emission: Fig.~\ref{beitraege}f and g)) is essential, as can be most 
clearly seen in case of $z_2$ ($^1S_0\to ^3P_0s$) and $z_3$ ($^1D_2\to ^3P_2s$). 
Especially, only after inclusion of the $\Delta$ in the final 
state the former became larger than the latter.

In this context it is also important to note that both types of contributions,
the emission of a real pion from a $\Delta$ decay---as depicted in diagrams d)
and f) of Fig. \ref{beitraege}---and the emission of a virtual pion from a
$\Delta$ that gets rescattered off the other nucleon---depicted in diagrams e)
and g) of Fig. \ref{beitraege}---are of similar numerical significance. This
should not come as a surprise, for as soon as the Delta--isobar is involved,
the large isovector pion nucleon interaction can contribute to the neutral
pion production \cite{juelichpol}. This is also consistent with the fact that
both these contributions (amongst others) contribute at next--to--leading
order in the chiral expansion \cite{withnorbert}.

In any case, it should be clear from this discussion that the
$\Delta$ degrees of freedom have to be taken into account
explicitly in any model that aims at a quantitative
description of the reaction $pp\to pp\pi^0$ even for energies near the
pion production threshold.

\section{Summary and outlook}

We have presented a partial-wave analysis of 
the double polarization data for the reaction $pp\to pp\pi^0$,
measured at the IUCF \cite{meyer}.  Due to
the limited statistical accuracy of the data, following the authors of
Ref. \cite{meyer}, we made several assumptions about the contributing
amplitudes in order to be able to perform the analysis.  The quality of the
 fit is with a $\chi^2$ per degree of freedom =1.7
not completely satisfying.  This could be a consequence of the several 
assumptions
that were made in the analysis.

  When compared to the results of a
microscopic model \cite{juelichdel}, the analysis
made three important points rather explicit: (i) the $\Delta$ degree of freedom
is important for a quantitative understanding of the reaction $pp\to pp\pi^0$,
(ii) there is especially one $z_\kappa$ that very strongly deviates from
that extracted from the data, namely $z_9$ ($^3P_1\to \, ^3P_0p$)---this will
guide the search for the possible short-comings of the model, and (iii) the 
set of partial waves included in the analysis was possibly too limited.

As a next major step a combined analysis of $NN$ scattering data and data on
$NN\to NN\pi$ needs to be performed. On the one hand, the pion production 
channels provide directly the inelasticities to be used for the analysis of the $NN$ 
data, on the other hand, the $NN$ elastic phase shifts provide the phase--motion as
well as the dominant $\epsilon$ dependence of the moduli of the production 
amplitudes.  The latter connection is provided by dispersion integrals as discussed in
detail in Refs. \cite{Goldbergerwatson,hanhart}.

\vspace{0.5cm}

\noindent 
{\bf Acknowledgments}

\noindent 
We would like to thank Ulf-G. Mei{\ss}ner for a careful reading of the
manuscript.  We also acknowledge communication with H.O. Meyer and L. Knutson.
P.N.D. acknowledges with thanks the support of the Alexander-von-Humboldt 
Foundation.

\end{document}